\documentclass[aps,prd,twocolumn,preprintnumbers,floatfix,nofootinbib]{revtex4-1}

\pdfoutput=1
\usepackage{amsmath}
\usepackage{amssymb}
\usepackage{amsfonts}
\usepackage{mathrsfs}
\usepackage{graphicx}
\usepackage{color}

\newcommand{\be}{\begin{equation}}
\newcommand{\ee}{\end{equation}}
\newcommand{\bea}{\begin{eqnarray}}
\newcommand{\eea}{\end{eqnarray}}

\newcommand{\df}{\dfrac}

\begin{document}

\preprint{IFT-UAM/CSIC-13-102 \\ $$}

\title{Quantifying the reheating temperature of the universe}
\author{Anupam Mazumdar$^{a}$}
\author{Bryan Zald\'\i var$^{b}$}
%\email{b.zaldivar.m@csic.es}

\vspace{1cm}
\affiliation{
${}^a$ Consortion for Fundamental Physics, Lancaster University, Lancaster,
LA1 4YB, UK \\
${}^b$Instituto de Fisica Teorica, IFT-UAM/CSIC, 
 28049 Madrid, Spain 
}

\begin{abstract}

The aim of this paper is to determine an exact definition of the reheat temperature for a generic perturbative decay of the inflaton.
In order to estimate the reheat temperature, there are two important conditions one needs to satisfy:  (a) the decay
products of the inflaton must dominate the energy density of the universe, i.e. the universe becomes completely radiation 
dominated, and  (b) the decay products of the inflaton have attained local thermodynamical equilibrium. For some choices of 
parameters, the latter is a more stringent condition, such that the decay products may thermalise much after the beginning of 
radiation-domination. Consequently, we have obtained that the reheat temperature can be much lower than the standard-lore estimation.  In 
this paper we describe under what conditions our universe could have {\it efficient} or {\it inefficient} thermalisation, and quantify the 
reheat temperature for both the scenarios. This result has an immediate impact on many applications which rely on the thermal 
history of the universe, in particular gravitino abundance.
\end{abstract}

\maketitle\

\section{introduction}

The transition from a cold inflating universe to a hot thermal universe depends {\it solely}  on the inflaton mass, $m_\phi$,  its coupling
$\alpha_\phi$ to the relevant degrees of freedom (d.o.f), and the dominant coupling between the decay products. In the case of Standard Model (SM)
particles, it is predominantly the strong interaction, $\alpha_s\sim 1/30$. This epoch is known as reheating~\cite{Albrecht:1982mp}, or preheating~\cite{preheating} (for a review see~\cite{Allahverdi:2010xz}). In this paper we will mostly concentrate on the case where the inflaton has a 
small Yukawa coupling to the relevant d.o.f., which would typically yield a perturbative decay of the inflaton to its almost massless quarks, leptons 
and gluons. This is well justified for a SM {\it gauge singlet} inflaton, since the SM quarks and leptons are {\it chiral} in nature, and therefore the 
lowest order couplings are determined by the dimensional $5$ operators in the potential, see~\cite{Mazumdar:2010sa}. Inflation could be driven 
by many independent sectors~\cite{Liddle:1998jc}, but what matters is the last field which is responsible for finally reheating the universe in our 
patch for the success of Big Bang Nucleosynthesis (BBN)~\cite{Beringer:1900zz}.

Especially, a SM gauge singlet inflaton could also couple to the SM Higgs with a 4-dimensional coupling, but through quartic 
coupling the inflaton never decays unless $\phi$ develops a VEV (vacuum expectation value):  
it rather leads to $\phi\phi \leftrightarrow HH$ scatterings, where $\phi$ is the inflaton and $H$ denotes the SM Higgs. In order 
to deplete the inflaton quanta it is still important to rely on the perturbative decay of the inflaton~\cite{GarciaBellido:2008ab}~\footnote{ Our 
treatment is very general and it can be applicable to supersymmetric theories. However there is a word of caution on how the inflaton couples to the 
supersymmetric Standard Model degrees of freedom, which depends very much on the origin of the inflaton. If inflaton is SM gauge singlet, see~\cite{Allahverdi:2007zz}, if inflaton is SM gauge invariant field, such as one belongs to the supersymmetric flat directions of squarks and 
sleptons~\cite{Enqvist:2003gh}, see  \cite{Allahverdi:2011aj}.}.

Typically, the reheating process is assumed to be instantaneous, with an efficient energy density conversion from the inflaton to the relativistic plasma. Within this framework the concept of reheating temperature $T_{rh}$ has been defined, see~\cite{Albrecht:1982mp,Kolb:1990vq}, ultimately relying on the assumption of the presence of local thermal equilibrium (LTE) at the very instant of conversion from the initial coherent oscillations of the inflaton domination to the radiation domination. 

The aim of this work is to determine a proper definition of the reheat temperature of the universe keeping in mind when the LTE is established along with the fact that 
the inflaton has completely decayed into radiation. When and how should we evaluate the reheat temperature is an important question for a number of applications ranging from evaluating the baryonic asymmetry, dark matter abundance and the 
success of BBN~\cite{Kolb:1990vq}. In this paper we shall put down the criteria of estimating the reheat temperature, based on when the inflaton decay products attain their thermalisation. Depending on whether the decay products of the inflaton thermalise before or after the radiation has dominated the universe, the reheat temperature will be very different. In either situation the notion of reheat temperature only makes sense when the universe is completely dominated by the radiation bath.

If thermalisation of the ambient plasma occurs during the coherent 
oscillations of the inflaton, one may be able to associate a maximum temperature for the relativistic species~\cite{Kolb:1990vq,Chung:1998zb}, but if the thermalisation time scale is longer than that of the inflaton-to-radiation domination transition time scale,  the notion of temperature does not make sense until the universe reaches its full LTE. In this respect there could be three regimes of 
interest which we will discuss in this paper:
\begin{enumerate}
\item{Instant thermalisation: when the inflaton decay products instantly thermalise upon decay.}
\item{Efficient thermalisation: when the inflaton decay products thermalise right at the instant when radiation epoch starts dominating the universe.}
\item{Delayed thermalisation: when the inflaton decay products thermalise deep inside the radiation dominated epoch after the transition from inflaton-to-radiation domination had occurred.}
\end{enumerate}

This paper is organised as follows. In section II we set the stage and write down the relevant equations for our analysis. The standard lore about the reheating epoch is briefly commented in section III. Section IV is devoted to present our analysis, in which we study the conditions under which the plasma attains thermalisation. Later on, in section V we discuss the concept of reheat temperature such as to properly capture the issues of thermalisation. Finally, we conclude in section VI. 

%%%%%%%%%%%%%%%%%%%%%%%%%%%%%%%%%

\section{Key assumptions and equations}

For the sake of simplicity, we will assume {\it universal} inflaton coupling, $\alpha_\phi$, to all its decay products, determined by
the number of relativistic d.o.f. $g_\ast$. Since the decay products of the inflaton are light, just from kinematics, they will typically have an initial  momentum roughly given by: $m_\phi/2$ for two-body decay, or $m_\phi/3$ for a three-body decay processes. 
The inflaton is assumed here to be a SM gauge singlet - it will decay universally to all its decay products, i.e. all the 
relativistic species $g_\ast$ would be excited.

Once the decay products are all excited there are two important processes which lead to thermalisation of all the d.o.f., or establish 
a LTE. Whereas a detailed thermalisation analysis of the plasma is out of the scope of this paper, some of its features are essential to our analysis, see Refs.~\cite{Enqvist:1990dp,Davidson:2000er}:

\begin{enumerate}

\item{\it Kinetic equilibrium}: Redistribution of the momentum between different decay particles. This can be achieved by number 
conserving $2\to 2$ scatterings with gauge boson exchange in the $t$-channel~\cite{Enqvist:1990dp,Davidson:2000er}. 

\item{\it Chemical equilibrium}: Number violating $2\to 3$ scatterings via $t$-channel 
are required to
establish the chemical equilibrium~\cite{Enqvist:1990dp,Davidson:2000er}. Higher order process are suppressed by further powers of the 
gauge coupling. Typically $2\to3$ interaction 
rate is higher than that of $2\to 2$. 

\end{enumerate}

The inelastic cross section for $2\to3$ processes are roughly estimated by~\cite{Davidson:2000er}:
\be
\sigma \sim \frac{\alpha_s^3}{p(t)^2}\log\left(\frac{m_\phi^2}{p(t)^2}\right)~,
\label{sigma}
\ee
where $\alpha_s\sim 1/30$ is the typical strong gauge coupling of the SM, and $p(t)$ is the 3-momentum  
transferred in the scattering process.

There are two interesting regimes which we will discuss below:
\begin{enumerate}
\item{{\bf $t$-channel enhancement}: If the scatterings $2\rightarrow 3$ processes via $t$-channel are mediated by light or massless gauge bosons, the cross section in question has an infrared divergence, which can be reasonably cut off by the Debye length, given by the inverse of the average separation between the two quanta, i.e. $\bar r\sim n^{-1/3}$, where $n$ is the number density of the particles in the plasma. In this case
the scattering rate is extremely fast due to the infrared divergence and would yield an efficient thermalisation of the plasma, as discussed in Refs.~\cite{Allahverdi:2002pu}.}
\item{{\bf $t$-chanel suppression}: As noted in \cite{Allahverdi:2002pu,Allahverdi:2007zz} this singularity is absent if, for example, the scattering happens via exchange of massive gauge boson. There, the thermalisation process may be considerably delayed due to suppression in the scattering rate. Such examples have been investigated in Refs.~\cite{Allahverdi:2007zz}, in presence of 
supersymmetric flat directions developing VEV or finite temperature effects, which naturally gives rise to massive gauge boson. In those cases, the most important processes for thermalisation are either $2\to3$ scatterings with scalar boson exchanges, or $s$-channel resonant gauge boson exchange. In either case, the infrared divergences disappear.}
\end{enumerate}
In this work we are going to discuss both the possibilities, although we will concentrate more on the delayed scenario, since the situation with enhanced cross sections has been extensively discussed in the literature \cite{Allahverdi:2002pu}.

On the other hand, the evolution of the inflaton, and the relativistic decay product's energy densities during the reheating period is described by the coupled set of Boltzmann equations, see~\cite{Kolb:1990vq}:
\bea
% \begin{displaymath}
 \bigg\{
 \begin{array}{ll}
\dot\rho_\phi + 3 H(t) \rho_\phi = -\Gamma_\phi \rho_\phi \\
\dot\rho_R + 4 H (t)\rho_R = \Gamma_\phi \rho_\phi  
+ \Gamma_{\rm th}(\rho_R -\rho_R^{\rm eq})~,
\label{Boltz1}
\end{array}
%\end{displaymath}
\eea
where the dots denote derivatives w.r.t. the physical time, $\rho_\phi (\rho_R)$ is the energy density of inflaton (radiation), being $\rho_R^{\rm eq}$ the equilibrium one; $H(t)$ is the Hubble parameter accounting for the expansion of the universe;  $\Gamma_\phi\equiv \alpha_\phi m_\phi$ is the inflaton decay rate\footnote{See \cite{Drewes:2013iaa} for a recent very detailed analysis on $\Gamma_\phi$.}, and $\Gamma_{\rm th}$  is the reaction rate responsible for thermalisation of the radiation plasma. Of course, once LTE is attained, the direct and inverse interactions among relativistic species counterbalance each other and the evolution of $\rho_R$ is dictated solely by the inflaton source and the Hubble expansion. 

%%%%%%%%%%%%%%%%%%%%%%%%%%%%%%%%%%%%%%%%%%%%%%%%%%%%%%%%%%%%

\section{Assuming LTE is established soon after inflaton decay}

Previous works which are relevant to our study have assumed LTE while studying the  evolution of the relativistic species during 
the reheating period~\cite{Kolb:1990vq}, see however \cite{Chung:1998zb,Enqvist:1990dp,Davidson:2000er,Allahverdi:2002pu} for emphasising the importance of acquiring LTE. At any epoch during reheating, as long as there is a relativistic bath in thermal equilibrium, we can extract an instantaneous temperature as:
\be
T(t) = \left[ \df{30}{\pi^2} \rho_R(t) / g_*(t)\right]^{1/4} 
\label{temp}
\ee
For a constant $g_*$ during the whole period, the evolution of the temperature according to Eq.~(\ref{temp}) is such that it has a maximum $T_{\rm max}$~\cite{Kolb:1990vq,Chung:1998zb}, which can be estimated as:
\be
T_{\rm max} \simeq \left[\df{1.57}{\pi^3 g_*}\right]^{1/4} \sqrt{M_P}~ (\Gamma_\phi H_I)^{1/4}
~,
\label{Tmax}
\ee
\newline
being $H_I$ the initial Hubble rate. Indeed, $T_{\rm max}$ can be potentially much larger than the reheating temperature, $T_{rh}$. The latter is usually defined as the temperature of the plasma
assuming an instantaneous conversion of the inflaton's energy density into radiation, 
at the time when $H(t) \approx \Gamma_\phi$,  such that:
\be
T_{rh}=\left(\df{90}{8\pi^3 g_*}\right)^{1/4} \sqrt{\Gamma_\phi M_P}~.
\label{Trh}
\ee

%%%%%%%%%%%%%%%%%%%%%%%%%%%%%%%%

\section{When is LTE  attained?}

However, LTE has to be attained and should not be taken for granted from the onset 
of the inflaton decay. In our analysis {\it we do not assume LTE} as a given condition for the relativistic species. 
Instead, we evaluate when and for which region of the inflaton parameters, $m_\phi$ and $\alpha_\phi$ for a fixed $\alpha_s=1/30$, this condition is achieved. 

There are (as justified later) two regions of the parameter space ($\alpha_\phi, m_\phi$), for which Eq. (\ref{Boltz1}) can be simplified such that the term 
$\Gamma_{\rm th}(\rho_R -\rho_R^{\rm eq})$ can be safely discarded: 
\begin{enumerate}
\item Very small $\alpha_\phi$ and very large $m_\phi$, for which $\Gamma_{\rm th}$ is very small:
\be
\Gamma_{\rm th} \ll \Gamma_\phi \cdot \left(\frac{\rho_\phi}{\rho_R}\right),~~~~\Gamma_{\rm th} \ll H
\ee
\item Very large $\alpha_\phi$ and very small $m_\phi$, for which $\rho_R\approx \rho_R^{\rm eq}$:
\be
\Gamma_{\rm th} \gg \Gamma_\phi \cdot \left(\frac{\rho_\phi}{\rho_R}\right),~~~~\Gamma_{\rm th} \gg H
\ee
\end{enumerate}
We will justify the notion of very small and very large  below. For these two cases, Eq.(\ref{Boltz1}) simplifies to 
(working with a comoving coordinate, $x\equiv a(t)\times m_\phi$, 
where $a(t)$ is the scale factor)~\cite{Chung:1998zb}:
\bea
 \left\{
 \begin{array}{cc}
\df{d\Phi}{dx} = -\left(\sqrt{\df{3}{8\pi}}\df{M_P}{m_\phi}\alpha_\phi \right)\df{x \Phi }{\sqrt{R+x\Phi}}\\
\df{dR}{dx} = \left(\sqrt{\df{3}{8\pi}}\df{M_P}{m_\phi}\alpha_\phi \right)\df{x^2 \Phi }{\sqrt{R+x\Phi}}
\label{Boltz1com}
\end{array}
\right.
\eea
with 
\be
\Phi\equiv \rho_\phi m_\phi^{-4} x^3,~
R\equiv \rho_R m_\phi^{-4} x^4~.
\ee
The initial condition is:
\be
R(x_I)=0,~~~\Phi_I \equiv \Phi(x_I) = \df{H_I^2 M_P^2}{8\pi/3}*m_\phi^{-4}x_I^3~,
\label{Phi0}
\ee
where the subindex $I$ refers to initial values. In many inflationary scenarios it is a good approximation to take $H_I\sim m_\phi$.

%%%%%%%%%%%%%%%%%%%%%%%%%%%%%%%%%%%%%%%%%%%%%%%%%%%%%%%
\begin{figure}[hbt]
\centering
\includegraphics[width=0.45\textwidth,angle=0]{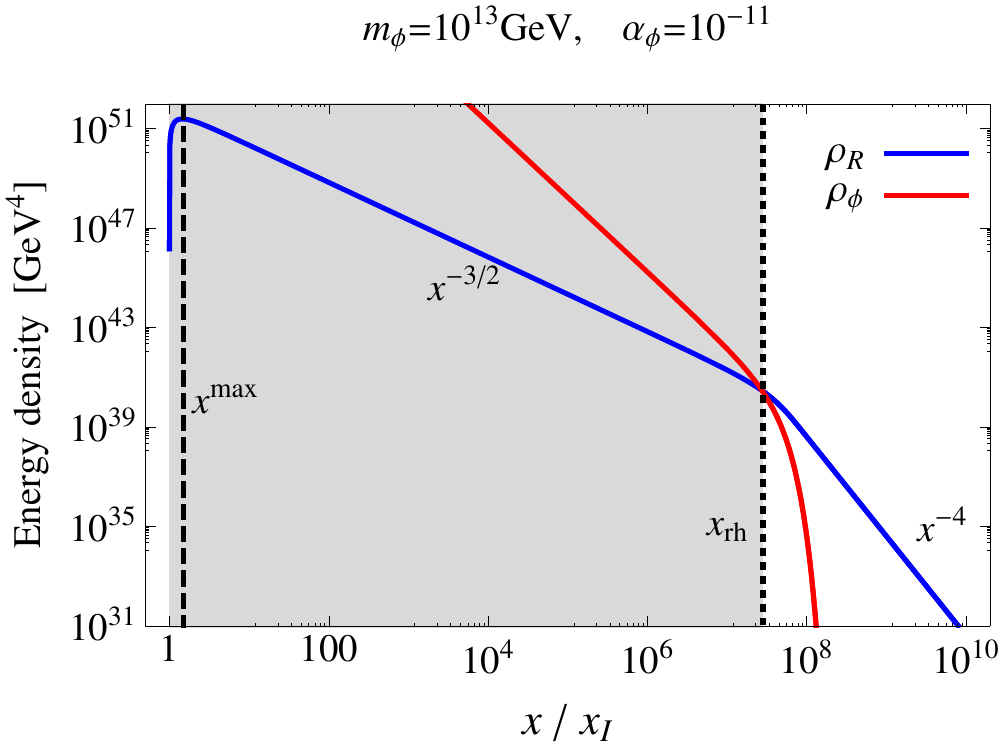}
\caption{\footnotesize{Radiation energy density (blue line), $\rho_R$, and inflaton energy density (red line), $\rho_\phi$, as a function of the scale factor, for $m_\phi=10^{13}$ GeV and $\alpha_{\phi}=10^{-11}$. The power laws indicate the behaviour of $\rho_R$ in the different regimes. The region in grey represents the reheating epoch, which by the {\it standard lore} finishes when radiation dominates the expansion (see text for details). }}
\label{fig:rhoR}
\end{figure} 
%%%%%%%%%%%%%%%%%%%%%%%%%%%%%%%%%%%%%%%%%%%%%%%%%%

We have solved Eq.~(\ref{Boltz1com}) numerically, and the result is shown in Fig.~\ref{fig:rhoR}, where for illustration we have taken $m_\phi=10^{13}$ GeV and $\alpha_{\phi}=10^{-11}$. We can infer that the radiation energy density (blue line) peaks very fast, around $x=x_{\rm max}\sim 1.5 x_I$, followed by a dilution due to the expansion. The position of the maximum is independent of the inflaton parameters. We also show for reference the inflaton energy density (red line), which as we can see completely dominates the expansion of the universe until the end of the reheating epoch. 
Analytically, during the inflaton-dominated period the radiation energy density goes like:
\bea
\rho^{id}_R(x) &\approx& \df{2}{5} \sqrt{\df{3}{8\pi}} \Gamma_\phi m_\phi^2 M_P \sqrt{\Phi_I}
~ x^{-3/2},~x_I\ll x< x_{rh} \nonumber \\
 &\approx& \df{0.15}{\pi} M_P^2 ~m_\phi^2\alpha_\phi \left(\df{x_I}{x}\right)^{3/2}
\label{rhoRsmall}
\eea
whereas for radiation-domination the expected $x^{-4}$-law is recovered:
\be
\rho^{rd}_R(x) \approx \rho^{id}_R(x_{rh}) \left(\frac{x_{rh}}{x}\right)^4,~~~~
x_{rh} < x~.
\label{rhoRlarge}
\ee
Here $x_{rh}$ (to be computed below) encodes the moment at which reheating ends. The super-indices ($id$) and ($rd$) stem for (inflaton-domination) and (radiation-domination), respectively.

The condition under which the plasma enters in thermal equilibrium can be naively estimated by the requirement 
\be
\Gamma_{\rm th} = n_R(x) \langle\sigma(x) v\rangle > H(x)~,
\label{Gamma}
\ee
where we approximate the cross-section $\sigma$ by Eq.~(\ref{sigma}), $v\approx c$ for relativistic species, and $n_R(x)$ is the relativistic number density. The latter can be directly extracted by solving Eq.~(\ref{Boltz1com}) in terms of number densities instead of energy densities. Assuming $2$-body decays of the inflaton (our results will not be affected much if we assume $3$-body decay of the inflaton), see also~\cite{Davidson:2000er}:
\be
n_R(x) \approx 2 n_\phi^I \left[
1-
e^{\left(-\Gamma_\phi\int_{x_0}^x \frac{d\tilde x}{\tilde x\cdot H(\tilde x)} \right) }
\right]
\left(\df{x_I}{x}\right)^3
\label{nR}
\ee
where the initial inflaton number density, $n^I_\phi \sim \rho^I_{\phi}/m_\phi$, as well as $H(x)$, are computed according to the our numerical solution of Eq.~(\ref{Boltz1com}).
Analytical estimations of Eq.~(\ref{nR}) can be obtained, as for the case of $\rho_R$, in two regimes. 

\begin{enumerate}
 
\item{ {\bf During inflaton-domination}: In this case, $R(x)$ gives a negligible contribution to the Hubble rate, whereas $\Phi$ remains approximately constant, $\Phi\approx\Phi_I$. In this case, it is straightforward to obtain:
\bea
n^{id}_R(x) &\simeq& 2n^I_\phi (1-e^{-\kappa x^{3/2}}) \left(\df{x_I}{x}\right)^3
\approx 2 n^I_\phi ~\kappa ~x_I^3 ~x^{-3/2} \nonumber\\
&\simeq& \df{0.5}{\pi} M_P^2 ~m_\phi \alpha_\phi \left(\df{x_I}{x}\right)^{3/2}~,
\label{nRsmall}
\eea
with $\kappa = (2/3)\alpha_\phi/x_I^{3/2}$. The superscript $'id'$ denotes {\it inflaton-domination}, since the inflaton oscillations are dominating over the relativistic species.}

\item{ {\bf During radiation-domination}: On the other hand, for {\it radiation-domination}, denoted below by the superscript $'rd'$, we clearly have:
\be
n^{rd}_R (x) \simeq 2n^I_\phi  \left(\df{x_I}{x}\right)^3
= \df{3}{4\pi} M_P^2~m_\phi \left(\df{x_I}{x}\right)^3~.
\label{nRlarge}
\ee
}
\end{enumerate}
The value $x_{rh}$ at which the regime changes could be computed in several ways, one of which is demanding $n_R^{id}(x_{rh}) = n_R^{rd}(x_{rh})$, resulting in:
\be
x_{rh} = \kappa^{-2/3} \simeq {1.3~ x_I}/{\alpha_\phi^{2/3}}.
\label{xrh}
\ee
Note that this value is independent of $m_\phi$ - heavier inflaton would have a shorter lifetime, but at the same time they would cause a faster expansion rates at early times.

%%%%%%%%%%%%%%%%%%%%%%%%%%%%%%%%%%%%%%%%%%%%%%%%%%%%%%%
\begin{figure}[hbt]
\centering
\includegraphics[width=0.45\textwidth,angle=0]{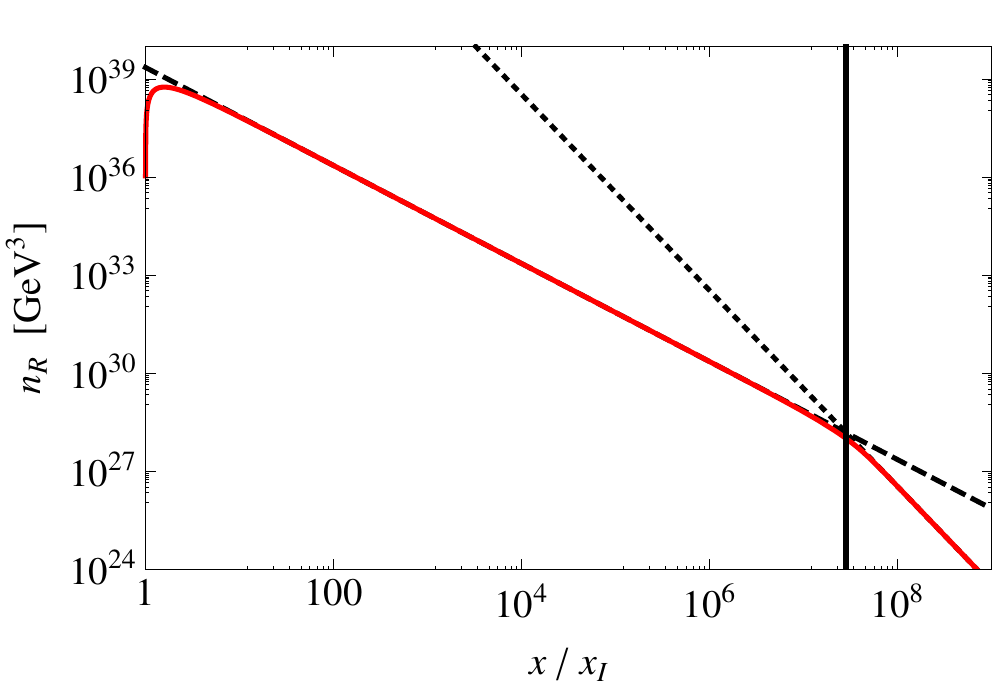}
\caption{\footnotesize{Radiation number density as a function of the scale factor, for $m_\phi=10^{13}$ GeV and $\alpha_{\phi}=10^{-11}$. The solid red line is the solution of Eq.~(\ref{nR}), where $H(x)$ is computed numerically from Eq.~(\ref{Boltz1com}). Dashed black line is the solution in Eq.~(\ref{nRsmall}), whereas the dotted black line is the solution in Eq.~(\ref{nRlarge}).  The solid black vertical lines is the value of $x_{rh}$ according to Eq.~(\ref{xrh}).}}
\label{fig:nR}
\end{figure} 
%%%%%%%%%%%%%%%%%%%%%%%%%%%%%%%%%%%%%%%%%%%%%%%%%%
We have shown in Fig.~\ref{fig:nR} the perfect agreement of the analytical estimations made in Eqs.~(\ref{nRsmall}-\ref{xrh}) w.r.t. the numerical solution in Eq.~(\ref{nR}). 

\subsection{Evolution of  the momenta of relativistic particles}
Coming back to the thermalisation analysis, since we cannot rely on an equilibrium distribution at this point, we take the typical momentum $\bar p(x)$ in Eq.~(\ref{sigma}) to be: 
\be
\bar p(x) = \df{d\rho_R(x)}{dn_R(x)} = \df{d\rho_R(x)}{dx}\cdot\left[\df{dn_R(x)}{dx}\right]^{-1}~.
\label{p(x)}
\ee
This expression directly follows from the definitions of $n_R$ and $\rho_R$, {\it without} assuming any particular shape of the distribution function $f(p)$. We would like to emphasise here that Eq.(\ref{temp}), in the absence of LTE, should not even have an interpretation of mean kinetic energy, since its functional shape incorporates the assumption of LTE-like $f(p)$.

%%%%%%%%%%%%%%%%%%%%%%%%%%%%%%%%%%%%%%%%%%%%%%%%%%%%%%%
\begin{figure}[hbt]
\centering
\includegraphics[width=0.45\textwidth,angle=0]{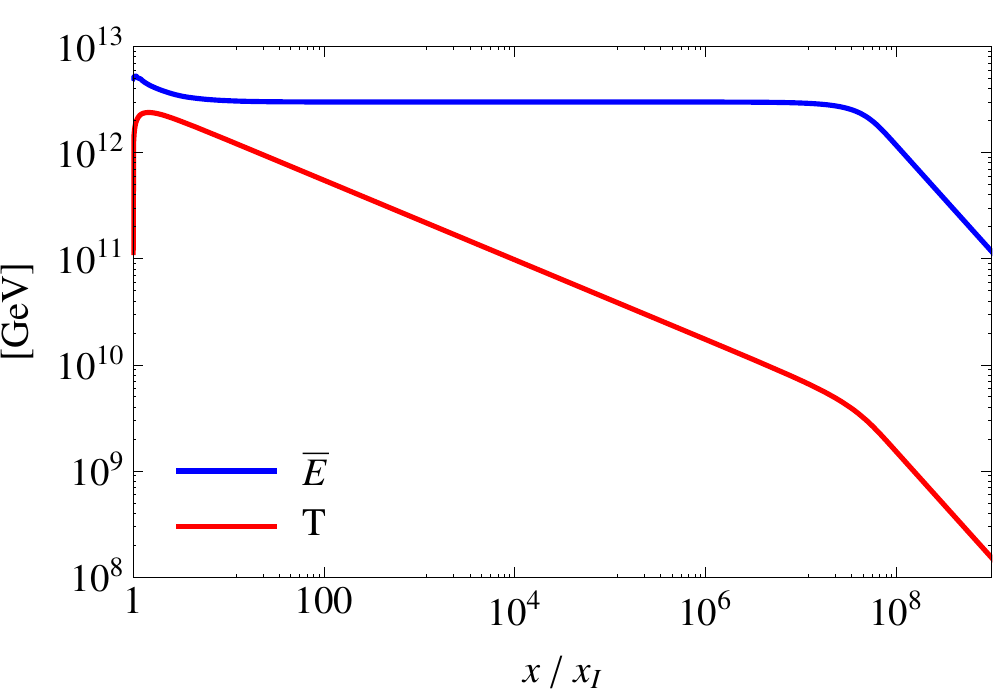}
\caption{\footnotesize{Comparison of the mean kinetic energy as computed according to Eq.~(\ref{p(x)}), and the temperature assuming LTE from the very onset of the inflaton decay as in Eq.~(\ref{temp}), for $m_{\phi}=10^{13}$ GeV, $\alpha_\phi=10^{-11}$.  One can see the obvious distinction and the importance of understanding when one should associate a temperature to the decay products of the inflaton.}}
\label{fig:temps}
\end{figure} 
%%%%%%%%%%%%%%%%%%%%%%%%%%%%%%%%%%%%%%%%%%%%%%%%%%

Taking then Eq.~(\ref{p(x)}) as a measure of the mean kinetic energy $\bar E$ of particles in the plasma, we compare $\bar E(x)$ with the temperature $T(x)$, extracted from Eq.~(\ref{temp}) under the assumption of thermal equilibrium. This is shown in Fig.~\ref{fig:temps}.
As can be observed, $\bar E$ is constant over almost the whole reheating period, 

whereas after reheating its evolution follows the same law as for $T(x)$, i.e. the well-known $T\propto x^{-1}$ behaviour of the radiation-dominated universe, resulting in:
\be
\bar E^{id} \approx m_\phi/3,~~~\bar E(x)^{rd} \simeq \df{0.5~ m_\phi}{\alpha_\phi^{2/3}}\df{x_I}{x}~.
\label{Elarge}
\ee
Physically it makes sense: during inflaton-domination the plasma (containing the relativistic species from the inflaton decay)  is getting constantly reheated
by the inflaton decay, and it turns out that it does so at a rate which is equal to the cooling rate due to the expansion. Afterwards, when the inflaton has decayed completely and only radiation remains, the energy of the relativistic species gets only redshifted by the expansion of the universe.

While this estimation for a typical momentum is reasonable in the scenario of delayed LTE, in the pure SM for example the emitted soft particles (out of the 2$\to$3 inelastic processes) may have momenta as low as:
\be
\bar p_{\rm cut} \sim n_R^{1/3}
\label{pcut}
\ee
where $n_R$ is given approximately by eq.(\ref{nRsmall}) or (\ref{nRlarge}) depending on the period of energy density domination. 
Comparing $\bar p_{\rm cut}$ wth $\bar E$ for the two regimes, we have that $\bar p_{\rm cut}< \bar E$ for:
\bea
&&m_\phi \gtrsim 2 M_P \alpha_\phi^{1/2} \left(\df{x_I}{x}\right)^{3/4},~~~~{(id-\rm epoch)} \\
&&m_\phi \gtrsim 1.4 M_P\alpha_\phi,~~~~{(rd-\rm epoch)}.\nonumber 
\eea
This means that in scenarios where the infrared enhancement is accessible, the thermalisation is much faster than in the delayed scenario, mainly for larger inflaton masses and smaller couplings. 

\subsection{Evaluating the thermalisation time}
As for the thermalisation condition is concerned, depending on the value of $(\alpha_\phi,m_\phi)$, in the delayed scenario the LTE  can be attained during inflaton-domination or afterwards, during radiation-domination. We should evaluate $\Gamma_{\rm th}$ by making use of the mean energies, $\bar E$, instead of the {\it temperature}, since as we pointed out above - we cannot rely at this point on thermal distribution. In the case of efficient thermalisation we use $\bar p_{\rm cut}$ instead. We then compare $\Gamma_{\rm th}$, according to the case, with:
\bea
H(x)\approx \left\{
\begin{array}{lr}
2.9~ m_\phi \left| \df{0.6x_I^{3/2}}{\sqrt{\pi}x^{3/2}} - 0.1\alpha_\phi\right|, & {(id-\rm epoch)} \\
~& ~\\
1.6\df{m_\phi}{\sqrt{\pi}\alpha_\phi^{1/3}}\df{x_I^2}{x^2}, & {(rd-\rm epoch)} 
\label{H}
\end{array}
\right.
 \eea
where in inflaton-domination , the Hubble rate is approximately given by: $H\propto(\rho_\phi^{id})^{1/2}$, whereas in radiation-domination case, we have: $H\propto(\rho_R^{rd})^{1/2}$. 

%%%%%%%%%%%%%%%%%%%%%%%%%%%%%%%%%%%%%%%%%%%%%%%%%%%%%%%
\begin{figure}[hbt]
\centering
\includegraphics[width=0.45\textwidth,angle=0]{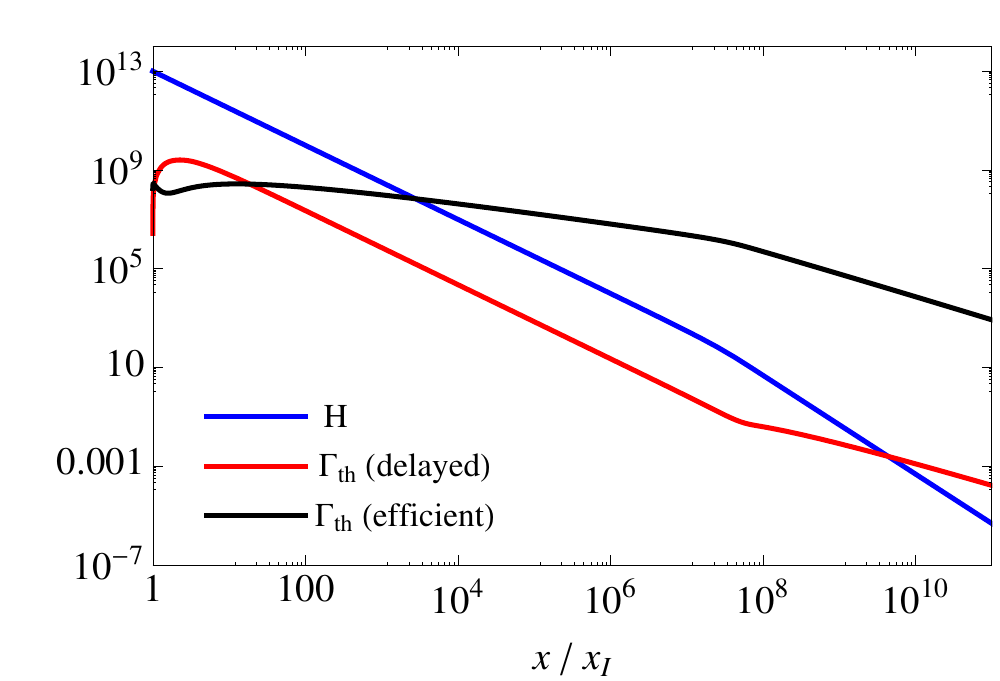}
\caption{\footnotesize{Reaction rate $\Gamma$ as a function of $x$ for the delayed (efficient) scenario in red (black), as compared to the Hubble parameter (blue). Break in the slopes determine the 
transition from inflaton domination to the radiation domination. Note that thermalisation time scale in the delayed scenario is larger than the matter-to-radiation transition scale.}}
\label{fig:thermal}
\end{figure} 
%%%%%%%%%%%%%%%%%%%%%%%%%%%%%%%%%%%%%%%%%%%%%%%%%%

We have shown in Fig.~\ref{fig:thermal} the comparison of $\Gamma_{\rm th}$ and $H$ from the numerical solution of Eq.~(\ref{Boltz1com}), using $\alpha_\phi=10^{-11}$ and $m_\phi=10^{13}$ GeV, for the sake of illustration, in the two scenarios: delayed and efficient thermalisation. 

For the efficient scenario, the thermalisation happens much before the beginning of radiation domination epoch, as expected.

On the other hand, for the delayed scenario, the evolution of $\Gamma_{\rm th}$ is parallel to that of $H$ for nearly the whole reheating period. Indeed,  in this region $\sigma(x)$ is nearly constant  (because $\bar E$ is) and thus $\Gamma_{\rm th}$ scales as $n_R(x)$, the latter evolving as $\rho_R(x)$ as was already deduced above (see Eqs.~(\ref{rhoRsmall}) and (\ref{nRsmall})). On the other hand the Hubble rate, even if dominated by the inflaton oscillations, also evolves as $\rho_R(x). $\footnote{This can be deduced from Eq.~(\ref{Phi0}) under the assumption of $\Phi\approx$const.} It is only after the inflaton population decreases substantially that the universe starts being radiation-dominated, thus the thermalisation processes become faster than the expansion rate and thermal equilibrium is achieved. The numerical solution for the thermalisation time, $x_{th}$, is around $x_{th} \sim10^{10}x_I$ for this choice of parameters.

Analytically it is possible to obtain the value of $x$ at which the thermalisation occurs, $\Gamma_{\rm th}(x_{th})=H(x_{th})$.  We just need to build up $\Gamma_{\rm th}$ from Eqs.~(\ref{nRlarge}) and (\ref{sigma}), whereas the Hubble rate is approximated by Eq.~(\ref{H}). As we are considering a delayed LTE scenario, we evaluate our cross section using (\ref{Elarge}) as explained above.

For the sake of illustration, assuming a total thermalisation cross-section which goes like $\sigma_{th} = \alpha_s^3/E^2$, see Eq.~(\ref{sigma}), we obtain the following solution for $x_{th}$:
\be
\frac{x_{th}^{rd}}{x_I} \approx \df{m_\phi^2}{\alpha_s^3~M_P^2 \alpha_\phi^{5/3}}
\label{xth}
\ee
for the case of a radiation-dominated thermalisation. However when including the log contribution, see Eq.~(\ref{sigma}), it is not possible to obtain an analytical solution of $x_{th}$. In this more accurate case, the solution is numerical and the thermalisation time $x_{th}$ is between 10 and 100 times smaller than what Eq.~(\ref{xth}) predicts.

On the other corner of the parameter space, for large $\alpha_\phi$ and small $m_\phi$, it usually happens that thermalisation happens very fast,  $x_{\rm th}^{id}\lesssim x_{\rm max}$,  when the inflaton still dominates the expansion. 

This is one of the main results of our analysis: for some choices of the pair $(\alpha_\phi,m_\phi)$, the plasma does not reach thermalisation at the time when the universe becomes radiation-dominated, but later. This happens for:
\be
\alpha_\phi \lesssim (0.01-0.1)\times \left(\frac{1 }
{\alpha_s^3}\right)\left(\frac{m_\phi}{M_p}\right)^2\,,
\label{lateth}
\ee 
where in the RHS we have corrected for the fact that a realistic $x_{th}$ may be $10^{-2}$ smaller than that of (\ref{xth}).

As an example for illustration, for a heavy mass, $m_\phi=10^{14}$ GeV, thermalisation reactions driven by $2\to 3$ processes of strong gauge coupling (as in 
Eq.~(\ref{sigma})),  the relativistic species  reaches thermal equilibrium later than the beginning of the radiation-domination era as long as 
$\alpha_{\phi} \lesssim 10^{-8}$.

Physically speaking this can be understood as follows:  even when the universe starts to become dominated by the radiation energy density, the thermalisation reaction rates may still be inefficient because of the very large typical energies of the interacting particles, $\bar E\lesssim {\cal O}(m_\phi)$, inherited from the inflaton decays and almost unaffected otherwise (see Fig.~\ref{fig:temps}, for an illustrative point). These large energies penalise the cross-sections, until the redshift is important enough as for the scattering process to become efficient enough, such that $\Gamma_{\rm th} > H$ and LTE is finally attained. Of course this works as long as the population of soft particles is not large enough as for affecting noticeably the rate of scattering processes.

%%%%%%%%%%%%%%%%%%%%%%%%%%%%%%%%%%%%%%

\section{Definition of reheat temperature}

Now let us define the reheating temperature, $T_{rh}$, as computed according to energy density (cf. Eq.\ref{temp}), provided the radiation have just thermalised, and  dominates the Hubble expansion rate of the universe. 
\be
T_{rh}= T(x),~~~~x={\rm max}(x_{th},x_{rh})~.
\label{defTrh}
\ee
There are three cases of interest:
\begin{enumerate}

\item{{\bf Instant thermalisation - $(x_{th}\ll x_{rh})$}:
Thermalisation of relativistic species is attained almost instantaneously (usually even around $x_{max}$), already during the coherent oscillations of the inflaton,
and they maintained LTE throughout reheating and also at the time when the universe becomes radiation dominated.
Following our prescription in Eq.~(\ref{defTrh}), in this case the reheat temperature is determined by: 
\be
T_{rh}(x_{th}\ll x_{rd}) \approx \df{0.6}{g_*^{1/4}} \sqrt{\alpha_\phi~m_\phi~M_P}~;
\label{TrhID}
\ee
A couple of points to note: we see that $T_{rh}(x_{th}\ll x_{rd})$ behaves exactly as the usual $T_{rh}$ of instant-reheating scenario, 
see Eq.~(\ref{Trh}), with an ${\mathcal O}(1)$-difference in a prefactor. Indeed  $T_{rh}(x_{th}< x_{rd})$ is a bit smaller than the usual 
definition of reheating case, as  assumed in Eq.~(\ref{Trh}), since the latter corresponds to a maximal thermalisation-efficiency by definition. In our case, the lower efficiency translates into a bit smaller reheating temperature (see Fig.~\ref{fig:Trh} below).  On the other hand, in this scenario, it is indeed possible to define a maximum temperature of the relativistic species, 
$T_{max} \equiv T(x_{max}) >T_{rh}$.
}

{\bf Efficient thermalisation}. This may be the case of the SM for example, where due to the presence of infrared divergences in eq.(\ref{sigma}), cured by a cut-off given in (\ref{pcut}), we get $x_{max}\ll x_{th}\leq x_{rh}$. In this case the estimation of $T_{rh}$ is again as in Eq.~(\ref{TrhID}) following the recipe Eq.~(\ref{defTrh}), and thermalisation happens within the inflaton-domination era.

\item{{\bf Delayed thermalisation - $(x_{th}\gg x_{rh})$}: Thermalisation happens deep inside the radiation dominated era, 
such that the reheat temperature is determined by:
\be
T_{rh}(x_{th}\gg x_{rd}) \approx (7-70)\times\df{\alpha_s^3~ \alpha_\phi^{3/2} M_P^{5/2}}{g_*^{1/4}m_\phi^{3/2}}~.
\label{TrhRD}
\ee
Note that  $T_{rh}(x_{th}\gg x_{rd})$ has an opposite behaviour with respect to $m_\phi$. This is the most important result of our work - in some region of the parameter space $(\alpha_\phi, m_\phi)$, where thermalisation happens after radiation starts dominating, the reheating temperature actually decreases with the inflaton mass.  Physically this is due to the following. For larger $m_\phi$, larger is the mean energy $\bar E$ of the relativistic species. This penalises the cross-sections for the thermalisation reactions which occur at the beginning, when soft processes are still unimportant, thus}  rendering the thermalisation rate less efficient at the end of the day, which is attained later. Consequently this lowers down  $T_{rh}(x_{th}\gg x_{rd})$.

\end{enumerate}
%%%%%%%%%%%%%%%%%%%%%%%%%%%%%%%%%%%%%%%%%%%%%%%%%%%%%%%
\begin{figure}[hbt]
\centering
\includegraphics[width=0.495\textwidth,angle=0]{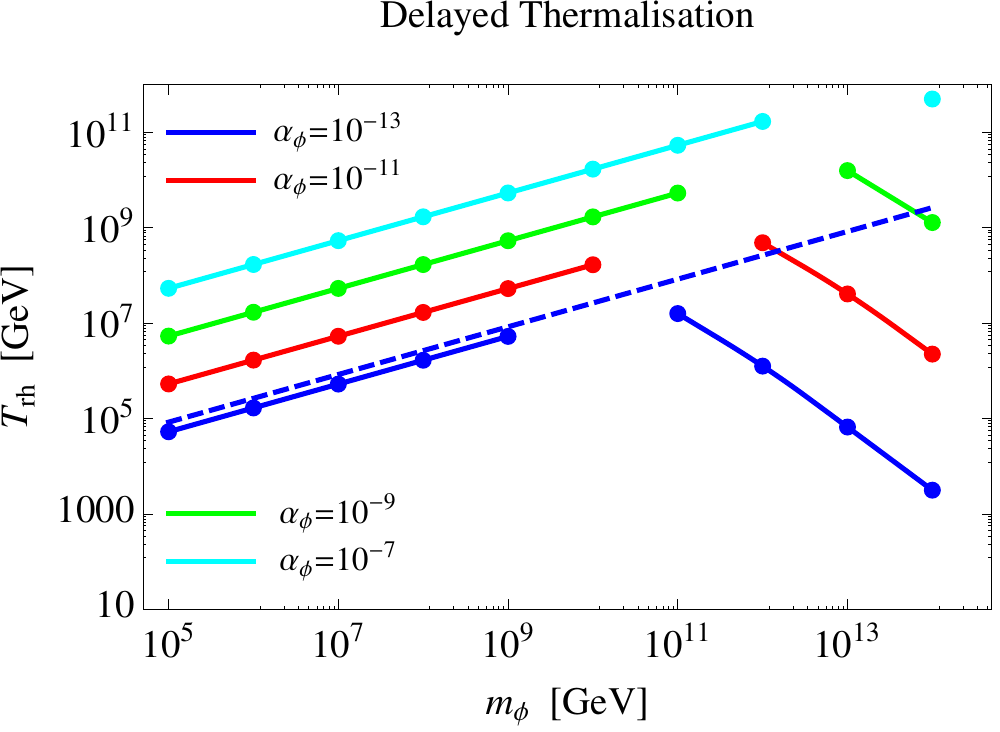}\\
\includegraphics[width=0.495\textwidth,angle=0]{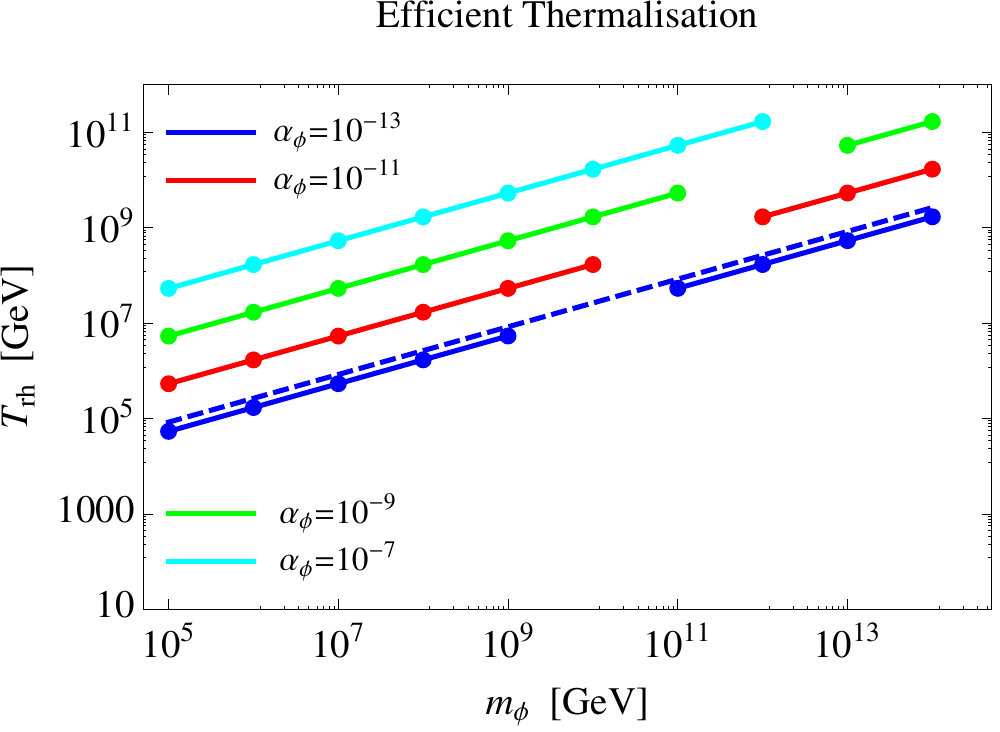}
\caption{\footnotesize{ Reheating temperature computed numerically by solving Eq.(\ref{Boltz1com}). This is represented with dots joined by solid lines for different values of $\alpha_\phi$: $10^{-13}$ (blue), $10^{-11}$ (red), $10^{-9}$ (green) and $10^{-7}$ (cyan), for $\alpha_s=1/30$. Delayed scenario is shown in top panel, whereas efficient scenario is shown in bottom panel. The blue dashed-line is an incorrect depiction of  reheating temperature (as in Eq.~(\ref{Trh})), corresponding to $\alpha_\phi = 10^{-13}$. }}
\label{fig:Trh}
\end{figure} 
%%%%%%%%%%%%%%%%%%%%%%%%%%%%%%%%%%%%%%%%%%%%%%%%%%

For scenarios beyond SM where large VEVs of scalar fields prevent the appearance of infrared divergences in (\ref{sigma})\footnote{i.e. where the ``efficient thermalisation" described above does not apply.} we have shown in Fig.~\ref{fig:Trh}-top the reheating temperature $T_{rh}$ as a function of $m_\phi$ for different values of $\alpha_\phi$, computed numerically by solving Eq.~(\ref{Boltz1com}) and represented with coloured dots joined by full lines.  As commented  above there are two regimes: one for which the thermalisation happens at inflaton-domination, where $T_{rh}(x_{th}< x_{rd})$ grows with $m_\phi$ and follows closely to 
$T_{rh}=({90}/{8\pi^3 g_*})^{1/4} \sqrt{\Gamma_\phi M_P}$ (see dashed blue line in Fig.~\ref{fig:Trh}); and a second regime for which the thermalisation happens deep inside radiation-domination era, where $T_{rh}(x_{th}> x_{rd})$ decreases with $m_\phi$.  Essentially, for the largest $\alpha_\phi$ and the smallest $m_\phi$, we are in the former regime, whereas for the smallest $\alpha_\phi$ and the largest $m_\phi$, we are in the latter regime. 

Note that, for example, for $\alpha_\phi=10^{-13}$ and $m_\phi=10^{13}$ GeV the usual $T_{rh}$ (as in Eq.(\ref{Trh})) largely overestimates the (more realistic) reheating temperature we have obtained in our analysis.  For the second regime, where Eq.~(\ref{TrhRD}) applies, we obtain a prediction in the correct ballpark for the numerical solution shown in Fig.~\ref{fig:Trh}. On the other hand the parametric dependence of Eq.~(\ref{TrhRD}) is verified.

A closer inspection of Fig.~\ref{fig:Trh} reveals some values of $m_\phi$ and $\alpha_\phi$ for which the numerical results are not shown. These ``holes" in the scan are due to the limited validity of our numerical solution of Eq.(\ref{Boltz1com}). As discussed above this expression, the $\Gamma_{\rm th}(\rho_R-\rho_R^{\rm eq})$ term is important when $\Gamma_{\rm th}$ becomes essentially comparable in size to the Hubble expansion and the inflaton source. Otherwise, either radiation-to-radiation terms are very inefficient (such that they do not play 
a role in the $\rho_R$-evolution), or if they are too efficient (such that production and annihilation balance each other in an equilibrium distribution), the 
Eq.~(\ref{Boltz1com}) is a reasonable simplification of the original Boltzmann set of equations Eq.~(\ref{Boltz1}). 

We have also obtained $T_{rh}$ numerically  in the ``efficient thermalisation" case, for all the parameter space (see Fig.~\ref{fig:Trh}-bottom). There, $T_{rh}$ grows monotonically as approximated by Eq.~(\ref{TrhID}) even for the largest inflaton masses, contrary to the delayed scenario where $T_{rh}$ decrease with $m_\phi$.

 \section{Phenomenological implications of delayed thermalisation}

The scenario of delayed thermalisation may have important implications for phenomenology and model building. We next briefly discuss some of the most direct ones:

\subsection{Leptogenesis}
 In this scenario \footnote{For a review see~\cite{Davidson:2008bu}.}, the existence of right-handed (RH) neutrinos give rise to the observed baryon asymmetry by sphaleron conversion processes. The lightest RH neutrino $N_1$ has to be massive enough as for producing sufficient CP asymmetry. Thus, since the thermal plasma needs to produce enough number density of those $N_1$ for a successful mechanism, a lower bound on the reheating temperature is imposed. Here we quote $T_{rh}\gtrsim 2\times10^9$ GeV   \cite{Rychkov:2007uq}. 
By direct inspection of Fig.~\ref{fig:Trh}, we can see that the leptogenesis bound forbids couplings $\alpha_\phi\lesssim 10^{-11}$ for whatever value of the inflaton mass. 

\subsection{Gravitinos}
Certainly one of the most important scenarios sensitive to the very early universe, the gravitino over-production poses a serious cosmological problem. The gravitinos are produced mostly at the reheating period out of the thermal bath and, if unstable, their late decays could potentially spoil the mechanisms leading to Big Bang Nucleosynthesis (BBN). On the other hand, if they are stable, they can over-close the Universe as dark matter candidates if the reheating temperature is large enough. Thus, in both cases we are able to place upper bounds on the reheating temperature.

\subsubsection{Unstable gravitinos}
\begin{figure}[hbt]
\centering
\includegraphics[width=0.495\textwidth,angle=0]{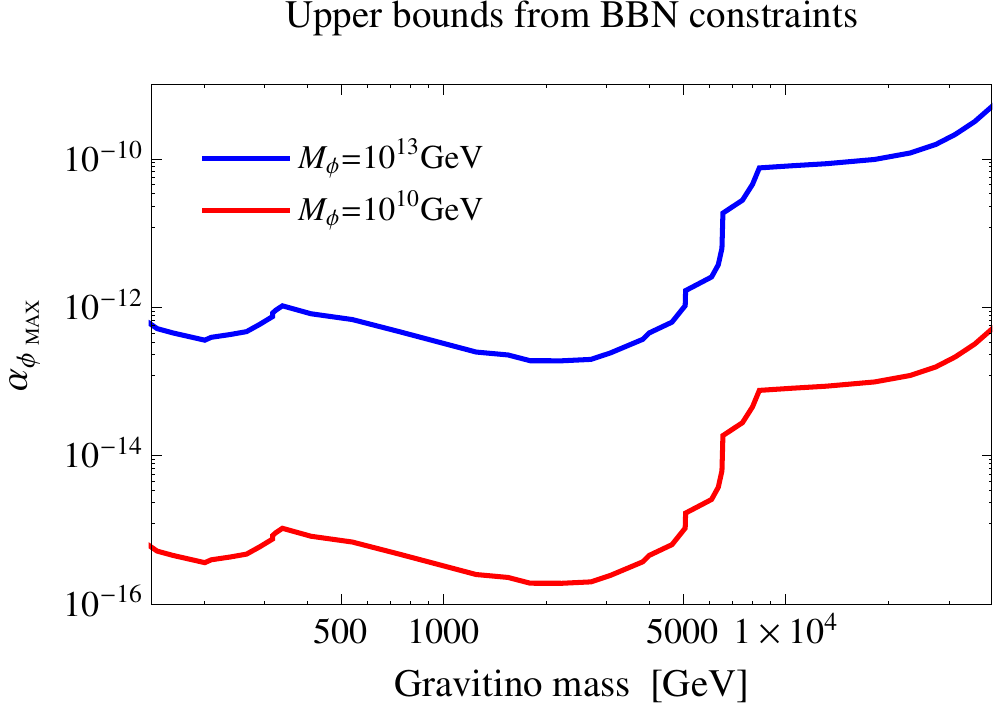}
\caption{\footnotesize{ Upper bounds on the coupling $\alpha_\phi$ coming from BBN constraints. This have been obtained by convoluting one of the bounds presented in \cite{Kawasaki:2008qe}. Upper (blue) line is for an inflaton mass of $M_\phi=10^{13}$ GeV, whereas the lower (red) line uses $M_\phi=10^{10}$ GeV.  }}
\label{fig:BBN}
\end{figure}

In this scenario the bounds on $T_{rh}$ come from the abundance of light elements. Given a situation of delayed thermalisation as the one we have discussed in this work, we could translate these upper bounds on $T_{rh}$ from BBN to bounds on the coupling $\alpha_\phi$ for  given inflaton masses,  as a function of the gravitino mass. The result is shown in Fig.~\ref{fig:BBN}. 
Based on \cite{Kawasaki:2008qe}, we have, for each gravitino mass $m_{3/2}$, maximum allowed values of $T_{rh}$ coming from the abundances of D, $^3$He, $^4$He, $^6$Li and $^7$Li elements. We take for each $m_{3/2}$ the strongest (i.e. the minimum) of these bounds, such that at the end of the day we have a convolution of these upper bounds as a function of $m_{3/2}$. Then, in Fig. \ref{fig:BBN} we have translated the obtained universal upper bound on $T_{rh}$, to upper bounds on $\alpha_\phi$, given two fixed inflaton masses: $10^{10}$ and $10^{13}$ GeV, by making use of our result  (\ref{TrhRD}). 

We see that in general the bounds from BBN greatly constrain the couplings $\alpha_\phi$: even for inflaton masses as large as $10^{13}$ GeV, couplings larger than $10^{-12}-10^{-10}$ are forbidden for a large range of gravitino masses. This bound goes in the opposite direction as the one from leptogenesis. Only if the gravitino is very heavy (larger than $10^4$ GeV) the two constraints are compatible, since then the BBN bound becomes loose.

%%%%%%%%%%%%%%%%%%%%%%%%%%%%%%%%%%%%%%%%%%%%%%%%%%%%%%%
%%%%%%%%%%%%%%%%%%%%%%%%%%%%%%%%%%%%%%%%%%%%%%%%%% 

\subsubsection{Stable gravitinos}
 
\begin{figure}[hbt]
\centering
\includegraphics[width=0.495\textwidth,angle=0]{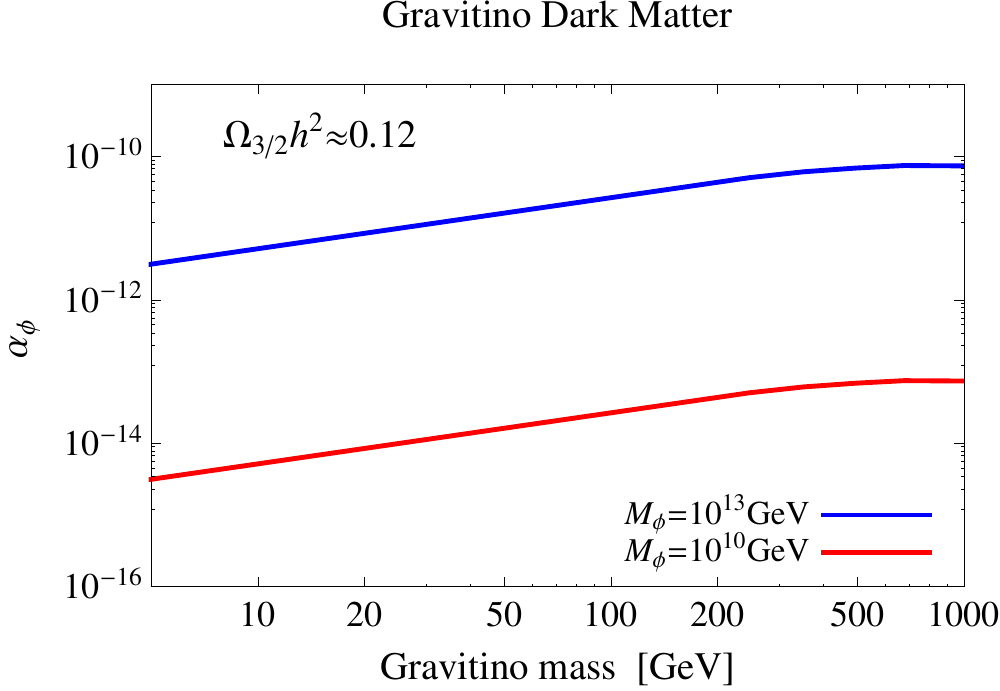}
\caption{\footnotesize{ Values of the coupling $\alpha_\phi$ for two inflaton masses which give rise to the gravitino relic abundance as full dark matter candidate. Colour code is the same as in previous figure. See text for more details. }}
\label{fig:gravitino}
\end{figure}

If the gravitinos were stable instead, thus dark matter candidates, their number density would freeze shortly after the reheating period. There are distinct ways in which gravitinos can be produced, for example, from direct perturbative decays of the inflaton~\cite{Maroto:1999ch}; from scatterings of the inflaton decay products (i.e.  relativistic species of supersymmetric Standard Model), or from thermal processes once the radiation bath has attained thermal equilibrium, see~\cite{Rychkov:2007uq}. However, as it is known the gravitino production before the thermalisation is attained gets diluted by the entropy release. Thus, it's final population is in very good approximation given by the thermal production yield, at the time where $T=T_{rh}$, the resulting relic abundance being \footnote{We have computed the evolution of the gravitino yield in this case. Our result is in agreement with \cite{Rychkov:2007uq}.}:
\be
\Omega_{3/2}h^2\approx 1.7\times10^{-3}\left(\df{m_{3/2}}{\rm GeV}\right)
\left(\df{T_{rh}}{10^{10}{\rm GeV}}\right)\left(\df{\gamma(T)}{T^6/M_P^2}\right)_{T=T_{rh}}
\ee  
 where $\gamma(T)$ is the total gravitino production rate, typically proportional to $T^6/M^2_P$, and dependent on the supersymmetric spectrum, particularly the gaugino masses. In fig. \ref{fig:gravitino} we show the prediction for a simplified scenario where the gaugino masses are degenerate and equal to 1 TeV at GUT scale. By fixing the gravitino relic abundance to $\Omega_{3/2}h^2=0.12$, in \cite{Rychkov:2007uq} they have obtained a prediction for $T_{rh}(m_{3/2})$, which we then translate to a prediction for the coupling $\alpha_\phi$ according to (\ref{TrhRD}), in a similar fashion as above. Amusingly, the resulting couplings are in the same ballpark as the upper bounds obtained above from a completely independent analysis.   
 
 %%%%%%%%%%%%%%%%%%%%%%%%%%%%%%%%%%%%%%%%%%%%%%%%%%%%%%%
%%%%%%%%%%%%%%%%%%%%%%%%%%%%%%%%%%%%%%%%%%%%%%%%%% 

 \section{Conclusions and discussions}

In this work we have studied inflationary reheating, in particular  revisiting the study of the  thermalisation of the inflaton decay products from both an analytical and numerical point of view, by analysing the dominant thermalisation process of the relativistic plasma as a whole. We have solved the coupled set of Boltzmann equations in two clearly defined regimes: a) The $2\to 3$ processes leading to thermalisation are too inefficient to affect the global evolution of the radiation energy-density itself, as a result the universe could be radiation dominated, but still not in local thermodynamical equilibrium (LTE),
and b) when thermalisation process is very quick,  at much larger rates compared to the Hubble expansion and the inflaton decay rate, in such a way that LTE of the decay products is attained very fast. In both regimes the Boltzmann equations are simplified in a similar fashion. 

We have obtained the following important results:
\begin{itemize}

\item  For sufficiently small $\alpha_\phi$ and sufficiently large inflaton-mass $m_\phi$, the relativistic plasma does not thermalises at the time where radiation-domination era begins, but (in some cases, much) later. When $\alpha_\phi$ is very small there are not enough relativistic species at the matter-to-radiation transition to immediately thermalise, whereas for very large $m_\phi$, the species are too energetic as for the relevant scattering processes to be efficient enough. Although, this requires significant suppression in $t$-channel scattering rate, which may happen for a massive gauge boson mediated interactions due to VEV or finite temperature effects giving mass to the gauge bosons. If there is a $t$-channel enhancement due to massless gauge boson mediation, then the scattering rate
is enhanced due to infrared effect and the Debye cut-off is determined by the number density of relativistic species present in the plasma. In this case thermalisation occurs during the inflaton oscillations 
dominating the universe.

\item We have determined a proper definition of the reheat temperature, in a generic scenario of perturbative decays of the inflaton. Essentially, two necessary conditions have to be eventually fulfilled: the plasma have to attain LTE, and it must dominate the expansion rate of the universe.  This is such that for some region of the inflaton parameters (precisely the one commented in the first point), the reheat temperature turns out to be much smaller than the standard estimations.  
\end{itemize}

Finally, we have discussed some connection with phenomenology by presenting implications on the gravitino cosmology. In general, for unstable or stable gravitinos, the predicted inflaton couplings need to be very small, order 
$\alpha_{\phi}\sim 10^{-12}-10^{-10}$ for an inflaton mass of $10^{13}$ GeV. 

Other phenomenological implications may be obtained in the context of Affleck-Dine baryogenesis (see e.g. ~\cite{Enqvist:2003gh}) and dark matter creation during reheating~\cite{Allahverdi:2002pu}. More recently  \ in the context of freeze-in mechanism through heavy portals \cite{Mambrini:2013iaa,Blennow:2013jba, Chu:2013jja}, we have other examples of DM which are sensitive to the reheat temperature. Some of this topical issues might as well have important implications for the inflaton mass and couplings, as for the gravitino case discussed above.

\section{Acknowldgements}
We thank E. Fernandez-Martinez and J. Rubio for very useful discussions. 
AM is supported by the Lancaster-Manchester-Sheffield Consortium for Fundamental Physics under STFC grant ST/J000418/1. 
BZ  acknowledges the Consolider-Ingenio PAU CSD2007-
00060, CPAN CSD2007-00042, under the contract FPA2010-17747; the Comunidad Autonoma de Madrid through the project
HEPHACOS P-ESP-00346, and the European Commission under contract PITN-GA-2009-237920, as well as the support of the Spanish MINECOÕs ÒCentro de Excelencia Severo OchoaÓ Programme under grant SEV-2012-0249.

%%%%%%%%%%%%%%%%%%%%%%%%%%%%%%%%%%%%%%%%%%%%%%%%%%


\begin{thebibliography}{99}

\bibitem{Albrecht:1982mp} 
  A.~Albrecht, P.~J.~Steinhardt, M.~S.~Turner and F.~Wilczek,
  %``Reheating an Inflationary Universe,''
  Phys.\ Rev.\ Lett.\  {\bf 48}, 1437 (1982).
  %%CITATION = PRLTA,48,1437;%%
  %307 citations counted in INSPIRE as of 09 Oct 2013
  
  \bibitem{preheating}
  J.~H.~Traschen and R.~H.~Brandenberger,
  %``Particle Production During Out-of-equilibrium Phase Transitions,''
  Phys.\ Rev.\ D {\bf 42}, 2491 (1990).
  %%CITATION = PHRVA,D42,2491;%%
    L.~Kofman, A.~D.~Linde and A.~A.~Starobinsky,
  %``Reheating after inflation,''
  Phys.\ RevY.~Shtanov, J.~H.~Traschen and R.~H.~Brandenberger,
  %``Universe reheating after inflation,''
  Phys.\ Rev.\ D {\bf 51}, 5438 (1995)
  [hep-ph/9407247].
  .\ Lett.\  {\bf 73}, 3195 (1994)
  [hep-th/9405187].
  L.~Kofman, A.~D.~Linde and A.~A.~Starobinsky,
  %``Towards the theory of reheating after inflation,''
  Phys.\ Rev.\ D {\bf 56}, 3258 (1997)
  [hep-ph/9704452].
  
\bibitem{Allahverdi:2010xz} 
  R.~Allahverdi, et.al,
  %``Reheating in Inflationary Cosmology: Theory and Applications,''
  Ann.\ Rev.\ Nucl.\ Part.\ Sci.\  {\bf 60}, 27 (2010)
  [arXiv:1001.2600 [hep-th]].
  %%CITATION = ARXIV:1001.2600;%%
  %71 citations counted in INSPIRE as of 09 Oct 2013

\bibitem{Mazumdar:2010sa} 
  A.~Mazumdar and J.~Rocher,
  %``Particle physics models of inflation and curvaton scenarios,''
  Phys.\ Rept.\  {\bf 497}, 85 (2011)
  [arXiv:1001.0993 [hep-ph]].
  %%CITATION = ARXIV:1001.0993;%%
  %147 citations counted in INSPIRE as of 10 Oct 2013


\bibitem{Liddle:1998jc} 
  A.~R.~Liddle, A.~Mazumdar and F.~E.~Schunck,
  %``Assisted inflation,''
  Phys.\ Rev.\ D {\bf 58}, 061301 (1998)
  [astro-ph/9804177].
  %%CITATION = ASTRO-PH/9804177;%%
  %283 citations counted in INSPIRE as of 14 Oct 2013


\bibitem{Beringer:1900zz} 
  J.~Beringer {\it et al.}  [Particle Data Group Collaboration],
  %``Review of Particle Physics (RPP),''
  Phys.\ Rev.\ D {\bf 86}, 010001 (2012).
  %%CITATION = PHRVA,D86,010001;%%
  %2404 citations counted in INSPIRE as of 14 Oct 2013



\bibitem{GarciaBellido:2008ab} 
  J.~Garcia-Bellido, D.~G.~Figueroa and J.~Rubio,
  %``Preheating in the Standard Model with the Higgs-Inflaton coupled to gravity,''
  Phys.\ Rev.\ D {\bf 79}, 063531 (2009)
  [arXiv:0812.4624 [hep-ph]].
  %%CITATION = ARXIV:0812.4624;%%
  %72 citations counted in INSPIRE as of 10 Oct 2013


\bibitem{Allahverdi:2007zz} 
  R.~Allahverdi and A.~Mazumdar,
  %``Reheating in supersymmetric high scale inflation,''
  Phys.\ Rev.\ D {\bf 76}, 103526 (2007)
  [hep-ph/0603244].
  %%CITATION = HEP-PH/0603244;%%
  %63 citations counted in INSPIRE as of 10 Oct 2013
   R.~Allahverdi and A.~Mazumdar,
  %``Supersymmetric thermalization and quasi-thermal universe: Consequences for gravitinos and leptogenesis,''
  JCAP {\bf 0610}, 008 (2006)
  [hep-ph/0512227].
  %%CITATION = HEP-PH/0512227;%%
  %64 citations counted in INSPIRE as of 10 Oct 2013


\bibitem{Enqvist:2003gh} 
  K.~Enqvist and A.~Mazumdar,
  %``Cosmological consequences of MSSM flat directions,''
  Phys.\ Rept.\  {\bf 380}, 99 (2003)
  [hep-ph/0209244].
  %%CITATION = HEP-PH/0209244;%%
  %225 citations counted in INSPIRE as of 10 Oct 2013
  
\bibitem{Allahverdi:2011aj} 
  R.~Allahverdi, et. al.
  %``Non-perturbative production of matter and rapid thermalization after MSSM inflation,''
  Phys.\ Rev.\ D {\bf 83}, 123507 (2011)
  [arXiv:1103.2123 [hep-ph]].
  %%CITATION = ARXIV:1103.2123;%%
  %25 citations counted in INSPIRE as of 10 Oct 2013

\bibitem{Kolb:1990vq} 
  E.~W.~Kolb and M.~S.~Turner,
  %``The Early Universe,''
  Front.\ Phys.\  {\bf 69}, 1 (1990).
  %%CITATION = FRPHA,69,1;%%
  %556 citations counted in INSPIRE as of 09 Oct 2013
  
  \bibitem{Chung:1998zb} 
  D.~J.~H.~Chung, E.~W.~Kolb and A.~Riotto,
  %``Production of massive particles during reheating,''
  Phys.\ Rev.\ D {\bf 60}, 063504 (1999)
  [hep-ph/9809453].
  %%CITATION = HEP-PH/9809453;%%
  D.~J.~H.~Chung, E.~W.~Kolb and A.~Riotto,
  %``Superheavy dark matter,''
  Phys.\ Rev.\ D {\bf 59}, 023501 (1999)
  [hep-ph/9802238].
  %%CITATION = HEP-PH/9802238;%%
  %220 citations counted in INSPIRE as of 09 Oct 2013

\bibitem{Enqvist:1990dp} 
  K.~Enqvist and K.~J.~Eskola,
  %``Thermalization in the Early Universe,''
  Mod.\ Phys.\ Lett.\ A {\bf 5}, 1919 (1990).
  %%CITATION = MPLAE,A5,1919;%%
  %13 citations counted in INSPIRE as of 09 Oct 2013
  K.~Enqvist and J.~Sirkka,
  %``Chemical equilibrium in QCD gas in the early universe,''
  Phys.\ Lett.\ B {\bf 314}, 298 (1993)
  [hep-ph/9304273].
  %%CITATION = HEP-PH/9304273;%%
  %22 citations counted in INSPIRE as of 09 Oct 2013

\bibitem{Davidson:2000er} 
  S.~Davidson and S.~Sarkar,
  %``Thermalization after inflation,''
  JHEP {\bf 0011}, 012 (2000)
  [hep-ph/0009078].
  %%CITATION = HEP-PH/0009078;%%
  %45 citations counted in INSPIRE as of 09 Oct 2013

\bibitem{Davidson:2008bu} 
  S.~Davidson, E.~Nardi and Y.~Nir,
  %``Leptogenesis,''
  Phys.\ Rept.\  {\bf 466}, 105 (2008)
  [arXiv:0802.2962 [hep-ph]].
  %%CITATION = ARXIV:0802.2962;%%
  %288 citations counted in INSPIRE as of 10 Oct 2013
  
  %\cite{Kawasaki:2008qe}
\bibitem{Kawasaki:2008qe}
  M.~Kawasaki, K.~Kohri, T.~Moroi and A.~Yotsuyanagi,
  %``Big-Bang Nucleosynthesis and Gravitino,''
  Phys.\ Rev.\ D {\bf 78} (2008) 065011
  [arXiv:0804.3745 [hep-ph]].
  %%CITATION = ARXIV:0804.3745;%%
  
  \bibitem{Rychkov:2007uq} 
  V.~S.~Rychkov and A.~Strumia,
  %``Thermal production of gravitinos,''
  Phys.\ Rev.\ D {\bf 75}, 075011 (2007)
  [hep-ph/0701104].
  %%CITATION = HEP-PH/0701104;%%

\bibitem{Allahverdi:2002pu} 
  R.~Allahverdi and M.~Drees,
  %``Thermalization after inflation and production of massive stable particles,''
  Phys.\ Rev.\ D {\bf 66}, 063513 (2002)
  [hep-ph/0205246].
   R.~Allahverdi and M.~Drees,
  %``Production of massive stable particles in inflaton decay,''
  Phys.\ Rev.\ Lett.\  {\bf 89}, 091302 (2002)
  [hep-ph/0203118].
  %%CITATION = HEP-PH/0203118;%%
  %50 citations counted in INSPIRE as of 28 Jun 2014
  
  
  \bibitem{Drewes:2013iaa}
  M.~Drewes and J.~U.~Kang,
  %``The Kinematics of Cosmic Reheating,''
  Nucl.\ Phys.\ B {\bf 875} (2013) 315
  [arXiv:1305.0267 [hep-ph]].
  %%CITATION = ARXIV:1305.0267;%%
  %6 citations counted in INSPIRE as of 06 Nov 2013


\bibitem{Maroto:1999ch} 
  A.~L.~Maroto and A.~Mazumdar,
  %``Production of spin 3/2 particles from vacuum fluctuations,''
  Phys.\ Rev.\ Lett.\  {\bf 84}, 1655 (2000)
  [hep-ph/9904206].
  %%CITATION = HEP-PH/9904206;%%
  %102 citations counted in INSPIRE as of 10 Jun 2014
  
%\cite{Mambrini:2013iaa}
\bibitem{Mambrini:2013iaa}
  Y.~Mambrini, K.~A.~Olive, J.~Quevillon and B.~Zaldivar,
  %``Gauge Coupling Unification and Non-Equilibrium Thermal Dark Matter,''
  Phys.\ Rev.\ Lett.\  {\bf 110} (2013) 241306
  [arXiv:1302.4438 [hep-ph]].
  %%CITATION = ARXIV:1302.4438;%%
  %3 citations counted in INSPIRE as of 13 Oct 2013

\bibitem{Blennow:2013jba}
  M.~Blennow, E.~Fernandez-Martinez and B.~Zaldivar,
  %``Freeze-in through portals,''
  arXiv:1309.7348 [hep-ph].
  %%CITATION = ARXIV:1309.7348;%%
  
  %\cite{Chu:2013jja}
\bibitem{Chu:2013jja}
  X.~Chu, Y.~Mambrini, JŽrŽm.~Quevillon and B.~Zaldivar,
  %``Thermal and non-thermal production of dark matter via Z'-portal(s),''
  arXiv:1306.4677 [hep-ph].
  %%CITATION = ARXIV:1306.4677;%%
  %3 citations counted in INSPIRE as of 13 Oct 2013

\end{thebibliography}
\end{document}